\documentclass[twocolumn,secnumarabic,amssymb,aps, pre]{revtex4}
\usepackage{amsmath}
\usepackage{graphicx}
\usepackage{float}

\usepackage[usenames,dvipsnames]{xcolor}
\usepackage{lipsum}

\begin{document}

\title{Tagged active particle: probability distribution in a slowly varying external 
potential is determined by effective temperature obtained from the Einstein relation}%

\author{Alireza Shakerpoor, Elijah Flenner and Grzegorz Szamel}%
\email[Email: ]{grzegorz.szamel@colostate.edu}
\affiliation{Department of Chemistry, Colorado State University, Fort Collins, Colorado 80523, USA}
\date{\today}
\begin{abstract}
We derive a distribution function for the position of a tagged active particle in a slowly varying in space external potential, 
in a system of interacting active particles. The tagged particle distribution has the form of the Boltzmann distribution 
but with an effective temperature that replaces the temperature of the heat bath. We show that the effective temperature that enters
the tagged particle distribution is the same as the effective temperature defined through the Einstein relation, 
\textit{i.e.} it is equal to the ratio of the self-diffusion and tagged particle mobility coefficients. 
This shows that this effective temperature, which is defined through a fluctuation-dissipation ratio, is 
relevant beyond the linear response regime. We verify our theoretical 
findings through computer simulations. Our theory fails when an additional large length scale appears in our active system. This length
scale is associated with long-wavelength density fluctuations that emerge upon approaching motility-induced phase separation. 
\end{abstract}
\maketitle

\section{Introduction}
	\label{sec:intro}

Equilibrium statistical mechanics provides us with explicit expressions for many-particle probability distributions for systems 
that are either isolated or in contact with one or more reservoirs~\cite{Chandler}. Probably the most often invoked 
distribution is the Boltzmann distribution $\propto\exp\left(-\mathcal{H}/T\right)$ describing an equilibrium system with Hamiltonian 
$\mathcal{H}$ at temperature $T$ (here and in the following we use units such that Boltzmann constant is equal to 1, $k_B=1$).
A lot of effort, analytical and/or numerical, is required to obtain from this distribution 
explicit results for measurable properties of a system of interacting particles,  
but at least we are provided with an explicit starting point for such an effort.

In contrast, for out-of-equilibrium stationary states we do not have such a starting point. If we were to follow the same
route as in equilibrium statistical mechanics, we would need to derive an exact or approximate expression for a non-equilibrium
steady-state many-particle distribution and then use it to calculate measurable properties of the non-equilibrium system considered. 
It is rather unlikely that a general formula for such a distribution exists. Conversely, it is very likely that if it were to be found
it would be more complicated than the many-particle equilibrium distribution. 

On the other hand, it is not clear that we need the full many-particle distribution. Most of the interesting properties 
of many-particle systems can be expressed in terms of reduced distribution functions, \textit{i.e.} pair distribution $g(r)$ and 
its generalizations to groups of more than two particles. To calculate these properties one can attempt to derive approximate 
formulas for the reduced distribution for specific non-equilibrium steady states. We note that in some cases the reduced distributions
in non-equilibrium steady states can be measured directly. For example, in the iconic scattering experiment of Clark and 
Ackerson~\cite{ClarkAckerson} the static structure factor, \textit{i.e.} the Fourier transform of the pair distribution function, 
of a sheared colloidal suspension was measured. This experiment inspired a number of other experimental, 
computational, and theoretical studies of the pair structure in colloidal systems under shear. 

In the present paper we focus on a class of non-equilibrium systems that have attracted a lot of attention in last decade, 
active matter systems
~\cite{Ramaswamy2010,Vicsek2012,Marchetti2013,Elgeti2015,Bechinger2016,Needleman2017,Ramaswamy2017}. 
The constituents of these systems consume energy and as a result move in a systematic way. Examples 
include assemblies of bacteria or of cells, suspensions of Janus colloidal particles, swarms of insects and flocks of birds. 
These constituents are often modeled as active or self-propelled particles, which move in a systematic way on short-time scales
and in a diffusive way on long-time scales. Importantly, their dynamics breaks detailed balance, and thus their stationary
states are profoundly different from equilibrium states. Needless to say, many-particle probability distributions describing 
these stationary states are not known explicitly. Several different approximate expressions for such distributions have been proposed 
and tested~\cite{Maggi2015,Farage2015,Rein2016}. 
In spite of a considerable body of work it is not yet clear which approximate method is most promising.

In some limits the problem of finding the many-particle stationary distribution for systems of 
interacting active particles may simplify. For example,
in a recent remarkable contribution de Pirey \textit{et al.}~\cite{dePirey2019} 
showed that in the large dimensional limit, higher-than-two-particle 
correlations are negligible and used this finding to derive an exact expression for the pair distribution function. 

Here we are interested in a more restricted problem. We consider a system of interacting active particles in the presence of
an external potential that varies slowly in space and acts on one particle only, the tagged particle. 
The question we want to answer is, what is the spatial distribution of tagged particle's position? 
For an equilibrium system at constant temperature this problem has a simple answer; the tagged particle distribution is
the Boltzmann distribution for a single particle in an external potential at the temperature of the system. Remarkably, this
answer is valid irrespectively of the spatial dependence of the external potential.   

We show that for a system of interacting active particles in the limit of slowly varying in space external potential the 
tagged particle distribution also has a form of the Boltzmann distribution. However, in this case the role of the temperature
is played by a variable that is a ratio of two quantities for which we derive exact albeit formal expressions. Importantly,
we show that these quantities are two well-known parameters describing tagged particle dynamics, the self-diffusion coefficient 
and the tagged particle mobility. Thus, the role of the temperature in our 
tagged particle distribution is played by the ratio of the self-diffusion and mobility coefficients, which has long been 
recognized as one of the so-called effective temperatures~\cite{Cugliandolo2011}, the Einstein relation temperature.

Recall that in equilibrium statistical mechanics the temperature appears not only in equilibrium
probability distributions but also in other relations. In particular, it appears  as a proportionality 
constant in fluctuation-dissipation relations, which connect fluctuations in equilibrium and linear response functions due to 
weak external perturbations~\cite{Chandler,Kubo1966,MarconiPuglisi2008}. The derivation of these relations relies upon the equilibrium
form of the many-particle distribution and in out-of-equilibrium systems these relations are generally not valid. 
In the nineties Cugliandolo, Kurchan and Peliti~\cite{Cugliandolo1997} realized that the violation of fluctuation-dissipation relations
can be used to define temperature-like quantities, which they called effective temperatures. 
These temperatures are defined through the fluctuation-dissipation ratios, \textit{i.e.} the ratios of the 
properties characterizing fluctuations and linear response/dissipation in non-equilibrium states. 
Importantly, Cugliandolo, Kurchan and Peliti 
showed that in a slowly relaxing  model system, the effective temperature determines the direction of the heat flow.   
Following this work, a number of different effective temperatures and their properties have been investigated in globally 
driven non-equilibrium stationary states~\cite{Berthier2002,Crisanti2003} and non-stationary aging systems~\cite{Barrat1999,Leuzzi2009}. 
Remarkably, in driven glassy systems it was found that several seemingly different temperatures have the 
same value~\cite{Berthier2002}, which hinted that there might be a unique effective temperature, at least in this case. 

More recently, the Einstein effective temperature, which is defined as  the ratio of the self-diffusion and tagged particle 
mobility coefficients, has been used to characterize some properties of active matter 
systems~\cite{Levis2015,Szamel2017,Petrelli2020,Loi2008}. In particular, some of us argued that 
the difference between the Einstein temperature and the so-called active temperature, which characterizes the strength of the 
self-propulsions, is a good measure of the departure of an active system from equilibrium~\cite{FlennerSzamel2020}. 

Since effective temperatures are defined through the ratio of fluctuations in a steady state to a function describing
linear response of this state to a weak external perturbation, it is not clear whether these temperatures can also describe
any non-linear response of steady states. Two studies showed the usefulness of the 
Einstein effective temperature for non-linear response. First, Hayashi and Sasa~\cite{Hayashi2004} showed that the Einstein
temperature determines the large scale distribution of a single Brownian particle moving in a tilted periodic potential.
Second, Szamel and Zhang~\cite{SzamelZhang2011} showed that the Einstein temperature determines the tagged particle density 
distribution in a slowly varying in space external potential in a system of interacting Brownian particles under steady shear.
In both cases the important assumption was the slow variation in space of the external potential, but there was no restriction on
its strength. 

The present result is similar to that of Ref.~\cite{SzamelZhang2011} in that we assume that the external potential acting 
on the tagged particle is slowly varying in space and we show that the density distribution is determined by the Einstein temperature. 
The important difference with this earlier work is in that the present system is athermal, locally driven by self-propulsions 
of individual particles, and isotropic.

We verify our theoretical results by performing computer simulations of an active system with an external potential. We show that the
the theory is valid as long as the spatial scale on which the tagged particle density distribution varies is the longest relevant 
length scale in the problem. 
When the density correlation length becomes large, due to the incipient motility-induced phase separation, the assumption 
behind our theory becomes invalid and numerical results show that the theory fails.

The paper is organized as follows. In Sec.~\ref{sec:theory} we present our theoretical derivation. In Sec.~\ref{sec:NumExp}, we 
describe our computer simulation model and describe our numerical procedures in Sec.~\ref{subsec:methods}, and then we
present the results and discuss the limitations of our theory  in Sec.~\ref{subsec:results}. Finally, we conclude the paper with 
an overview of our results in Sec.~\ref{sec:conclusions}.

\section{Theoretical Derivation}
	\label{sec:theory}

To derive the equation describing the tagged active particle density distribution in a slowly varying in space external potential
we use a gradient expansion. Specifically, we use a version of the celebrated Chapman-Enskog expansion that was originally introduced 
to derive hydrodynamic equations and the expressions for transport coefficients from the 
Boltzmann kinetic equation~\cite{Resibois}. The specific implementation of the Chapman-Enskog
procedure that we use is inspired by Titulaer's \cite{Titulaer1978} derivation of the generalized Smoluchowski equation, which describes 
diffusive motion of a colloidal particle in an external potential, from the Fokker-Planck equation, 
which describes the motion of the same particle on a shorter
time scale, using both particle's positions and momentum. Our present derivation is similar to that used earlier \cite{SzamelZhang2011}
to obtain an equation describing the tagged particle distribution in a sheared colloidal suspension. 

To make the derivation concrete we need to specify the active particles model. We consider a system of active Ornstein-Uhlenbeck
particles (AOUPs) \cite{Szamel2014,Maggi2015,Fodor2016}. These particles move  
in a viscous medium, without inertia, under the combined influence of the inter-particle 
forces and self-propulsions, with the latter evolving according to
the Ornstein-Uhlenbeck stochastic process.
The equations of motions read
\begin{eqnarray}
\label{eq:motion1}
\dot{\mathbf{r}}_i &=& \mu_0\left[\mathbf{F}_i + \mathbf{f}_i\right], \\
\tau_p \dot{\mathbf{f}}_i &=& \mathbf{f}_i + \boldsymbol{\eta}_i.
\label{eq:motion2}
\end{eqnarray}
In Eq.~\eqref{eq:motion1} $\mathbf{r}_i$ is the position of particle $i$,
$\mu_0$ is the mobility coefficient of an isolated particle, which is the inverse of the isolated particle's friction coefficient,
$\mu_0=\xi_0^{-1}$, $\mathbf{F}_i$ is
the force acting on particle $i$ due to all other particles, 
\begin{equation}\label{eq:force}
\mathbf{F}_i=\sum_{j\neq i} \mathbf{F}(\mathbf{r}_{ij}),
\end{equation}
where $\mathbf{r}_{ij}=\mathbf{r}_i-\mathbf{r}_j$ and $\mathbf{F}(\mathbf{r})=-\partial_{\mathbf{r}}V(r)$ with
$V(r)$ being the two-body potential, 
and $\mathbf{f}_i$ is the self-propulsion.
In Eq.~\eqref{eq:motion2} $\tau_p$ is the persistence time
of the self-propulsion and $\boldsymbol{\eta}_i$ is the internal Gaussian noise
with zero mean and variance
$\left< \boldsymbol{\eta}_i(t) \boldsymbol{\eta}_j(t^\prime) \right>_{\mathrm{noise}}
= 2 \xi_0 k_B T_a \mathbf{I} \delta_{ij} \delta(t-t^\prime)$,
where $\left< \ldots \right>_\mathrm{noise}$ denotes averaging over the noise
distribution, $T_a$ is the ``active'' temperature, and $\mathbf{I}$ is the unit tensor. The active temperature characterizes
the strength of the self-propulsion. In addition, it determines the long-time diffusion coefficient of an isolated AOUP,
$D_0=T_a\mu_0\equiv T_a/\xi_0$. 

We assume that there is a slowly varying in space external potential, $\Phi(\mathbf{r}_1)$,
acting on particle 1. This particle will be referred to as the tagged particle. The external potential results in an additional term,
$- \partial_{\mathbf{r}_1} \Phi(\mathbf{r}_1)$, in the equation of motion for the tagged particle,
\begin{eqnarray}
\label{eq:motion3}
\dot{\mathbf{r}}_1 &=& \mu_0\left[\mathbf{F}_1 - \partial_{\mathbf{r}_1} \Phi(\mathbf{r}_1) + \mathbf{f}_1\right], \\
\tau_p \dot{\mathbf{f}}_1 &=& \mathbf{f}_1 + \boldsymbol{\eta}_1.
\label{eq:motion4}
\end{eqnarray}

We assume that the systems described by equations of motion (\ref{eq:motion1}-\ref{eq:motion3}) can reach a stationary state.
The $N$-particle stationary state distribution of positions and self-propulsions, $P_{\mathrm{s}}^\Phi$,
satisfies the following equation,
\begin{equation}\label{eq:PsF}
\left[\Omega_s + \partial_{\mathbf{r}_1}\cdot\mu_0\left(\partial_{\mathbf{r}_1} \Phi(\mathbf{r}_1)\right)\right]
P_{\mathrm{s}}^\Phi\left(\mathbf{r}_{1},\mathbf{f}_{1},\ldots, \mathbf{r}_{N},\mathbf{f}_{N}\right)=0.
\end{equation}
Here $\Omega_s$ is the evolution operator that corresponds to the unperturbed equations of motion,
\begin{multline}
\Omega_{\mathrm{s}}\left(\mathbf{r}_{1},\mathbf{f}_{1},\ldots, \mathbf{r}_{N},\mathbf{f}_{N}\right)=
-\mu_0 \sum_{i=1}^{N} \partial_{\mathbf{r}_{i}} \cdot\left(\mathbf{F}_i+\mathbf{f}_{i}\right)
 \\
+\sum_{i=1}^{N} \partial_{\mathbf{f}_{i}} \cdot\left(\frac{1}{\tau_{p}}\mathbf{f}_{i} 
+\frac{T_{\text{a}}}{\mu_0\tau_{p}^{2}} \partial_{\mathbf{f}_{i}}\right).
	\label{eq:EvolutionOperator}
\end{multline}

To make the assumption that the external potential acting on the tagged particle explicit, we write it as 
$\Phi(\epsilon \mathbf{r}_1)$, where $\epsilon$ is a small parameter. As described before \cite{SzamelZhang2011},
we will use $\epsilon$ as an expansion parameter and then, at the end of the derivation, we will set it to 1.

Our goal is to derive from Eq. (\ref{eq:PsF}) a closed equation for the stationary tagged particle density distribution, 
$n_s(\mathbf{r}_1)$,
\begin{equation}
n_s(\mathbf{r}_1) = \int d\mathbf{f}_1 d\mathbf{r}_2 \ldots d\mathbf{f}_N d\mathbf{r}_N
P_{\mathrm{s}}^\Phi\left(\mathbf{r}_{1},\mathbf{f}_{1},\ldots, \mathbf{r}_{N},\mathbf{f}_{N}\right).
\end{equation}
The tagged particle density is non-uniform due to the external potential $\Phi$. Due to the slow
variation of the external potential, we assume that
the tagged particle density will also be slowly varying. Again, to make this assumption explicit
we write the tagged particle density as $n_s(\epsilon \mathbf{r}_1)$.

Due to the inter-particle interactions the $N$-particle distribution is not a slowly varying function of the tagged 
particle position, if the positions of all other particles are kept constant. However, it should be a slowly varying function
of $\mathbf{r}_1$ if it is written in terms of the tagged particle position and positions of all other particles \emph{relative}
to the tagged particle position, \textit{i.e.} in terms of $\mathbf{r}_1$ and $\mathbf{r}_{21}$, $\mathbf{r}_{31}$ \textit{etc.} 
To make this assumption explicit we change the variables and write the stationary state equation in terms of 
$\mathbf{R}_{1}=\epsilon\mathbf{r}_{1}$, $\mathbf{R}_{2}=\mathbf{r}_{21}$, \ldots, $\mathbf{R}_{N}=\mathbf{r}_{N1}$,
\begin{widetext}
\begin{equation}\label{eq:PsFexp}
\left[\Omega_s^{(0)} + \epsilon \Omega_s^{(1)} 
+ \epsilon^2 \partial_{\mathbf{R}_1}\cdot\mu_0\left(\partial_{\mathbf{R}_1} \Phi(\mathbf{R}_1)\right)\right]
P_{\mathrm{s}}^\Phi\left(\mathbf{R}_{1},\mathbf{f}_{1},\ldots, \mathbf{R}_{N},\mathbf{f}_{N}\right)=0,
\end{equation}
where we separate contributions to the evolution operator of different orders in $\epsilon$,
\begin{align} \label{eq:OmegaExpansion0}
&\Omega^{(0)}=-\mu_0\left[-\sum_{i=2}^{N} \partial_{\mathbf{R}_{i}} 
\cdot\left(\sum_{i\neq j=2}^N \mathbf{F}\left(-\mathbf{R}_{j}\right)+\mathbf{f}_{1}\right)
+\sum_{i=2}^{N} \partial_{\mathbf{R}_{i}} 
\cdot\left(\sum_{i \neq j=2}^N \mathbf{F}\left(\mathbf{R}_{i}\right)+\mathbf{f}_{i}\right)\right]
+\sum_{i=1}^{N} \partial_{\mathbf{f}_{i}} \cdot\left(\frac{1}{\tau_{p}} \mathbf{f}_{i}+\frac{T_{\text{a}} }{\mu_0\tau_{p}^{2}}
\partial_{\mathbf{f}_{i}}\right), 
\\
&\Omega^{(1)} =-\mu_0 \partial_{\mathbf{R}_{1}} 
\cdot\left(\sum_{i=2}^N \mathbf{F}\left(-\mathbf{R}_{i}\right)+\mathbf{f}_{1}\right)
-\sum_{i=2}^{N} \partial_{\mathbf{R}_{i}} \cdot \mu_0 \left(\partial_{\mathbf{R}_{1}} \Phi\left(\mathbf{R}_{1}\right)\right).
 	\label{eq:OmegaExpansion1}
\end{align}
\end{widetext}

Following Titulaer \cite{Titulaer1978} and Ref. \cite{SzamelZhang2011}, we now look for a special perturbative solution 
of Eq. (\ref{eq:PsFexp}) 
\begin{multline}
P_{\mathrm{s}}^\Phi\left(\mathbf{R}_{1},\mathbf{f}_{1},\ldots, \mathbf{R}_{N},\mathbf{f}_{N}\right)=
 \\
n_{\mathrm{s}}\left(\mathbf{R}_{1}\right) 
P_{\mathrm{s}}^{(0)}\left(\mathbf{f}_{1},\mathbf{R}_{2},\mathbf{f}_{2}, \ldots, \mathbf{R}_{N},\mathbf{f}_{N}\right)
 \\
+\epsilon P_{\mathrm{s}}^{(1)}\left(\mathbf{R}_{1},\mathbf{f}_{1},\mathbf{R}_{2},\mathbf{f}_{2}, \ldots, 
\mathbf{R}_{N},\mathbf{f}_{N}\right)
 \\
+\epsilon^{2} P_{\mathrm{s}}^{(2)}\left(\mathbf{R}_{1},\mathbf{f}_{1},\mathbf{R}_{2},\mathbf{f}_{2}, \ldots, 
\mathbf{R}_{N},\mathbf{f}_{N}\right)+\ldots.
	\label{eq:ProbDensityExpansion}
\end{multline}
We use the solution postulated in Eq. (\ref{eq:ProbDensityExpansion}) to derive perturbatively 
an equation for the tagged particle density distribution,
\begin{equation}
\left(\mathcal{D}^{(0)}+\epsilon \mathcal{D}^{(1)}+\epsilon^{2} \mathcal{D}^{(2)}+\ldots\right) 
n_{\mathrm{s}} \left(\mathbf{R}_{1}\right)=0.
	\label{eq:TaggedFokkerPlanckExpansion}
\end{equation}
Eq. (\ref{eq:TaggedFokkerPlanckExpansion}) is obtained by the integration of Eq. (\ref{eq:PsFexp}) over 
the self-propulsion of the tagged particle and the positions of all particles other than the tagged particle. For example, the
first two terms in Eq. (\ref{eq:TaggedFokkerPlanckExpansion}) read
\begin{align} \label{eq:D0ns}
& \mathcal{D}^{(0)} n_{\mathrm{s}} \left(\mathbf{R}_{1}\right) =
\int d \mathbf{f}_{1} d \mathbf{R}_{2} d \mathbf{f}_{2} \ldots d \mathbf{R}_{N} d \mathbf{f}_{N} 
\Omega^{(0)} n_{\mathrm{s}} \left(\mathbf{R}_{1}\right) P_{\mathrm{s}}^{(0)},
 \\
& \mathcal{D}^{(1)} n_{\mathrm{s}} \left(\mathbf{R}_{1}\right) = \int d \mathbf{f}_{1} d \mathbf{R}_{2} d \mathbf{f}_{2} \ldots
d \mathbf{R}_{N} d \mathbf{f}_{N} \Omega^{(1)} n_{\mathrm{s}} \left(\mathbf{R}_{1}\right) P_{\mathrm{s}}^{(0)} \nonumber
 \\
& + \int d \mathbf{f}_{1} d \mathbf{R}_{2} d \mathbf{f}_{2} \ldots d \mathbf{R}_{N} 
d \mathbf{f}_{N} \ \Omega^{(0)} P_{\mathrm{s}}^{(1)} .
\label{eq:D1ns}
\end{align}
We note that the second term in Eq. (\ref{eq:D1ns}) vanishes due to integration by parts. 

Following the standard Chapman-Enskog procedure \cite{Resibois,Titulaer1978}, the tagged particle density 
$n_{\mathrm{s}}\left(\mathbf{R}_{1}\right)$ is not expanded in $\epsilon$. Moreover, as in the standard 
Chapman-Enskog procedure \cite{Resibois,Titulaer1978}, there is some freedom in choosing higher order functions $P_{\mathrm{s}}^{(i)}$,
$i\ge 1$. This freedom is eliminated  by imposing the usual conditions,
\begin{equation}\label{eq:Psifix}
\int d \mathbf{f}_{1} d \mathbf{R}_{2} d \mathbf{f}_{2} \ldots d \mathbf{R}_{N} d \mathbf{f}_{N} 
P_{\mathrm{s}}^{(i)}=0 \hspace{20pt} \forall \hspace{5pt} i\ge 1.
\end{equation}
Conditions (\ref{eq:Psifix}) imply that the tagged particle density is completely determined by the zeroth order term
in expansion (\ref{eq:PsFexp}).

To find the special solution for the stationary state probability distribution we substitute (\ref{eq:ProbDensityExpansion}) into 
Eq. (\ref{eq:PsFexp}) and solve order by order. The terms of zeroth order give
\begin{equation}\label{eq:PsFexp0th}
\Omega^{(0)} P_{\mathrm{s}}^{(0)}\left(\mathbf{f}_{1},\mathbf{R}_{2},\mathbf{f}_{2}, \ldots, \mathbf{R}_{N},\mathbf{f}_{N}\right) =0.
\end{equation}
Thus, $P_{\mathrm{s}}^{(0)}$ is the translationally invariant steady state distribution of the positions and self-propulsions
in the absence of the external potential. The combination of the expansion (\ref{eq:ProbDensityExpansion}) and conditions 
(\ref{eq:Psifix}) implies that this distribution should be normalized to 1,
\begin{equation}\label{eq:Ps0fix}
\int d \mathbf{f}_{1} d \mathbf{R}_{2} d \mathbf{f}_{2} \ldots d \mathbf{R}_{N} d \mathbf{f}_{N} 
P_{\mathrm{s}}^{(0)}=1.
\end{equation}

The first order terms give an equation for $P_{\mathrm{s}}^{(1)}$ in which $n_{\mathrm{s}}P_{\mathrm{s}}^{(0)}$ plays 
the role of a source term,
\begin{equation}\label{eq:PsFexp1st}
\Omega^{(0)} P_{\mathrm{s}}^{(1)} + \Omega^{(1)} n_{\mathrm{s}}\left(\mathbf{R}_{1}\right) P_{\mathrm{s}}^{(0)}=0.
\end{equation}
We can formally solve Eq. (\ref{eq:PsFexp1st}) for $P_{\mathrm{s}}^{(1)}$, 
\begin{equation}
P_{\mathrm{s}}^{(1)} = -  \left[\Omega^{(0)}\right]^{-1} \Omega^{(1)} n_{\mathrm{s}}\left(\mathbf{R}_{1}\right) P_{\mathrm{s}}^{(0)}.
\end{equation}
We recall that $P_{\mathrm{s}}^{(0)}$ does not depend on $\mathbf{R}_1$ and we get
\begin{multline}\label{eq:P1P0}
P_{\mathrm{s}}^{(1)} =  \mu_0 \left[\Omega^{(0)}\right]^{-1} 
\left[\sum_{i=2}^{N} \mathbf{F}\left(-\mathbf{R}_{i}\right)+\mathbf{f}_{1}\right]P_{\mathrm{s}}^{(0)}\cdot
\partial_{\mathbf{R}_{1}} n_{\mathrm{s}} \left(\mathbf{R}_{1}\right)
\\
+ \mu_0 \left[\Omega^{(0)}\right]^{-1} \left(\sum_{i=2}^{N} \partial_{\mathbf{R}_{i}}\right)P_{\mathrm{s}}^{(0)}\cdot
\left(\partial_{\mathbf{R}_{1}} \Phi \left(\mathbf{R}_{1}\right)\right) n_{\mathrm{s}}\left(\mathbf{R}_{1}\right).
\end{multline}

We then use these results to derive successive terms in the stationary state equation for the tagged particle distribution, 
Eq. (\ref{eq:TaggedFokkerPlanckExpansion}). We note that $\mathcal{D}^{(0)}$, Eq. (\ref{eq:D0ns}), 
involves $\Omega^{(0)}P_{\mathrm{s}}^{(0)}$, and thus
it vanishes. Then, we note that $\mathcal{D}^{(1)}$, Eq. (\ref{eq:D1ns}), consists of two terms and, as we stated earlier, that
the second term vanishes due to integration by parts. In turn, the first term, involving $\Omega^{(1)}$, consists of
two contributions that originate  from the 
two contributions to $\Omega^{(1)}$, Eq. (\ref{eq:OmegaExpansion1}). 
The first one is proportional to the following integral,
\begin{multline}\label{eq:taggedcurrent}
\mu_0 \int d \mathbf{f}_{1} d \mathbf{R}_{2} d \mathbf{f}_{2} \ldots
d \mathbf{R}_{N} d \mathbf{f}_{N} \left(\sum_{i=2}^N \mathbf{F}\left(-\mathbf{R}_{i}\right)+\mathbf{f}_{1}\right)
P_{\mathrm{s}}^{(0)},
\end{multline}
which involves the sum of the total inter-particle force acting on that tagged particle and of the self-propulsion of the tagged particle.
We note that integral (\ref{eq:taggedcurrent}) is equal to the tagged particle current in the unperturbed stationary state. 
We assume that there are no average stationary currents in the stationary state, and thus integral (\ref{eq:taggedcurrent}) vanishes.
The term contributing to $\mathcal{D}^{(1)}$ that originates from the second contribution to $\Omega^{(1)}$, 
Eq. (\ref{eq:OmegaExpansion1}), vanishes due to integration by parts.

The lowest order non-vanishing contribution to stationary state equation (\ref{eq:TaggedFokkerPlanckExpansion}) originates
from $\mathcal{D}^{(2)}$,
\begin{multline}\label{eq:D2ns}
\mathcal{D}^{(2)} n_{\mathrm{s}} \left(\mathbf{R}_{1}\right) = 
\\
\int d \mathbf{f}_{1} d \mathbf{R}_{2} d \mathbf{f}_{2} \ldots
d \mathbf{R}_{N} d \mathbf{f}_{N} 
\partial_{\mathbf{R}_1}\cdot\mu_0\partial_{\mathbf{R}_1} \Phi(\mathbf{R}_1)n_{\mathrm{s}} \left(\mathbf{R}_{1}\right) P_{\mathrm{s}}^{(0)}
\\
+ \int d \mathbf{f}_{1} d \mathbf{R}_{2} d \mathbf{f}_{2} \ldots
d \mathbf{R}_{N} d \mathbf{f}_{N} 
\Omega^{(1)} P_{\mathrm{s}}^{(1)}
\\ + \int d \mathbf{f}_{1} d \mathbf{R}_{2} d \mathbf{f}_{2} \ldots d \mathbf{R}_{N} 
d \mathbf{f}_{N} \ \Omega^{(0)} P_{\mathrm{s}}^{(2)} .
\end{multline}
The first term at the right-hand-side gives 
$\partial_{\mathbf{R}_1}\cdot\mu_0\partial_{\mathbf{R}_1} \Phi(\mathbf{R}_1)n_{\mathrm{s}} \left(\mathbf{R}_{1}\right)$ and 
the last term vanishes after integration by parts. The second term is a sum of two contributions that originate from the 
two contributions to $\Omega^{(1)}$, Eq. (\ref{eq:OmegaExpansion1}). The second one vanishes after integration by parts and the 
first one can be re-written as
\begin{equation}\label{eq:D2ns2}
\partial_{\mathbf{R}_{1}} D \cdot \partial_{\mathbf{R}_{1}} n_{\mathrm{s}}\left(\mathbf{R}_{1}\right) + 
\partial_{\mathbf{R}_{1}} \bar{\mu} \cdot \left(\partial_{\mathbf{R}_{1}} \Phi \left(\mathbf{R}_{1}\right)\right)
n_{\mathrm{s}}\left(\mathbf{R}_{1}\right)
\end{equation}
where
\begin{align}
& D = -\frac{\mu_0^2}{d} \int d \mathbf{f}_{1} d \mathbf{R}_{2} d \mathbf{f}_{2} \ldots
d \mathbf{R}_{N} d \mathbf{f}_{N} \left(\sum_{i=2}^N \mathbf{F}\left(-\mathbf{R}_{i}\right)+\mathbf{f}_{1}\right)\cdot
\nonumber \\ 
& \times 
\left[\Omega^{(0)}\right]^{-1} 
\left[\sum_{i=2}^{N} \mathbf{F}\left(-\mathbf{R}_{i}\right)+\mathbf{f}_{1}\right]P_{\mathrm{s}}^{(0)},
\label{eq:D}
\\ \nonumber 
& \bar{\mu} = - \frac{\mu_0^2}{d} \int d \mathbf{f}_{1} d \mathbf{R}_{2} d \mathbf{f}_{2} \ldots
d \mathbf{R}_{N} d \mathbf{f}_{N} \left(\sum_{i=2}^N \mathbf{F}\left(-\mathbf{R}_{i}\right)+\mathbf{f}_{1}\right)\cdot
\\ & \times 
\left[\Omega^{(0)}\right]^{-1} \left(\sum_{i=2}^{N} \partial_{\mathbf{R}_{i}}\right)P_{\mathrm{s}}^{(0)}.
\label{eq:deltamu}
\end{align}
We note that while writing Eqs. (\ref{eq:D}-\ref{eq:deltamu}) we used the rotational invariance of the $d$-dimensional 
stationary state without the external potential.

Combining all non-vanishing contributions to $\mathcal{D}^{(2)}$, setting $\epsilon=1$, 
and re-writing the resulting stationary state equation in terms of the original 
coordinate $\mathbf{r}_1$ we get the following equation for the tagged particle distribution, 
\begin{equation}\label{eq:nsfinal}
\partial_{\mathbf{r}_{1}} \cdot \left[ D \partial_{\mathbf{r}_{1}}  + 
\mu \left(\partial_{\mathbf{r}_{1}} \Phi \left(\mathbf{r}_{1}\right)\right)\right] n_{\mathrm{s}}\left(\mathbf{r}_{1}\right) =0
\end{equation}
where, rewritten in terms or the original coordinates $\mathbf{r}_1, \ldots, \mathbf{r}_N$, $D$ and $\mu$ read
\begin{eqnarray}
D &=& -\frac{\mu_0^2}{d} \frac{1}{V} 
\int \mathrm{d} \mathbf{r}_{1} \mathrm{d} \mathbf{f}_{1} \ldots \mathrm{d} \mathbf{r}_{N} \mathrm{d} \mathbf{f}_{N} 
\left(\mathbf{F}_{1}+\mathbf{f}_{1}\right)\cdot
\nonumber \\ && \times \left[\Omega_{\mathrm{s}}\right]^{-1}\left(\mathbf{F}_{1}+\mathbf{f}_{1}\right) 
P_{\mathrm{s}}^{(0)}\left(\mathbf{r}_{1},\mathbf{f}_{1}, \ldots, \mathbf{r}_{N},\mathbf{f}_{N}\right)
\label{eq:Dfinal}
\\  
\mu &=& \mu_0 - \frac{\mu_0^2}{d} \frac{1}{V} 
\int \mathrm{d} \mathbf{r}_{1} \mathrm{d} \mathbf{f}_{1} \ldots \mathrm{d} \mathbf{r}_{N} \mathrm{d} \mathbf{f}_{N} 
\left(\mathbf{F}_{1}+\mathbf{f}_{1}\right)\cdot
\nonumber \\ && \times \left[\Omega_{\mathrm{s}}\right]^{-1}
\partial_{\mathbf{r}_{1}} P_{\mathrm{s}}^{(0)}\left(\mathbf{r}_{1},\mathbf{f}_{1}, \ldots, \mathbf{r}_{N},\mathbf{f}_{N}\right)
\label{eq:mufinal}
\end{eqnarray}

Eq. (\ref{eq:nsfinal}) implies that the tagged particle distribution has the Boltzmann form,
\begin{equation}\label{eq:TaggedProbDens}
n_{\mathrm{s}}\left(\mathbf{r}_{1}\right) \propto 
\exp \left(-\Phi\left(\mathbf{r}_{1}\right) / T^{\text {eff}}\right),
\end{equation}
where the effective temperature is the ratio of $D$ and $\mu$,
\begin{equation}\label{eq:EffTemp}
T^{\text{eff}}=D/\mu.
\end{equation}

The final step in the derivation is to assign some physical interpretation to $D$ and $\mu$. This interpretation has already been 
hinted by our choice of the symbols we used for these quantities. 
First, we note that since $\mu_0\left(\mathbf{F}_1+\mathbf{f}_1\right)$ is the tagged particle velocity, 
$D$ can be formally interpreted as the integral of the velocity
auto-correlation function,
\begin{equation}\label{eq:DVACF}
D = d^{-1} \int_0^\infty \left< \dot{\mathbf{r}}_1(t)\cdot  \dot{\mathbf{r}}_1(0)\right>,
\end{equation}
which in turn is the standard expression for the self-diffusion coefficient \cite{Chandler}. 
Second, we note that if, for a system initially in a stationary state, a weak spatially uniform external force $\mathbf{F}_1^\text{ext}$
is applied to the tagged particle, the tagged particle will start moving. Initially, since the distribution of the other particles 
around the tagged particle is isotropic, its velocity will be equal to $\mu_0 \mathbf{F}_1^\text{ext}$ but after some time the 
the distribution of the other particles will become slightly anisotropic and they will be exerting an additional friction force on the
tagged particle. It can be shown that due to the change of the probability distribution the long-time limit of the tagged particle 
velocity will be $\left(\mu_0 + \bar{\mu}\right)\mathbf{F}_1^\text{ext}$. Thus, $\mu=\mu_0 + \bar{\mu}$, Eq. (\ref{eq:mufinal}), 
is the tagged particle mobility coefficient.

To summarize, we showed in this section that for a slowly varying in space external potential acting on the tagged particle 
the tagged particle density has the Boltzmann form with the temperature determined by the ratio of the self-diffusion and
mobility coefficients, \textit{i.e.} the Einstein effective temperature.  

\section{Numerical Verification}
	\label{sec:NumExp}
\subsection{Methods}
	\label{subsec:methods}
	
To test Eq. (\ref{eq:TaggedProbDens}) for the probability distribution and Eq. (\ref{eq:EffTemp}) for the effective temperature, 
we performed a series of computer simulations of interacting AOUPs in an external potential $\Phi$, evolving according to equations
of motion (\ref{eq:motion1}-\ref{eq:motion4}), and a parallel series of computer simulations of unperturbed particles,
evolving according to equations of motion (\ref{eq:motion1}-\ref{eq:motion2}). We simulated $N=10^4$ particles 
at a number density of $\rho=0.6751$. 
We used the finite range Weeks-Chandler-Andersen (WCA) purely repulsive pair potential 
$V^{\text{WCA}}(r)=4\varepsilon\left[\left(\frac{\sigma}{r}\right)^{12} - \left(\frac{\sigma}{r}\right)^{6}\right]+\varepsilon$, 
for $r<2^{1/6}\sigma$ and zero otherwise.
We present the results in standard LJ units where $\varepsilon$ is the unit of energy, $\sigma$ is the unit of length, and 
$\sigma^2/(\mu_0 \varepsilon)$ is the unit of time.

We simulated AOUP systems at two different active temperatures, $T_a=0.01$ and $T=1.0$ and a range of persistence times $\tau_p$.
The values of $T_a$ were chosen to roughly represent two different dependencies of the self-diffusion coefficient on the persistence
time, which we identified in an earlier investigation \cite{FlennerSzamel2020,Berthier2017}. 
At the lower temperature we expected $D$ either to increase with $\tau_p$
or to have a non-monotonic dependence on $\tau_p$. In contrast, at the higher temperature we expected $D$ 
to decrease with $\tau_p$.

For simulations without the external potential we used time step $dt=0.001$ for $\tau_p\ge 0.02$ and $dt=0.0002$ for $\tau_p=0.002$.
For the perturbed systems we used $dt=0.001$ for $T_a=0.01$ (at which $\tau_p\ge 0.02$) and $dt=0.0001$ for $T_a=1.0$.

\subsubsection{Evaluation and analysis of the tagged particle distribution}
	\label{subsubsec:T_eff}

To induce a slowly varying, non-uniform density distribution of the tagged particle we used a potential that is periodic over the 
simulation box length $L$, and varying along the $x$, $y$ or $z$ axis,
\begin{equation}
\Phi(\alpha) = \Phi_{0} \sin\left( 2\pi\alpha/L \right) \hspace{20pt} \alpha=x,y,z.
	\label{eq:ExtPotential}
\end{equation}
Since for the $N=10^4$ particle system $L$ is much larger than the particle size, this potential is indeed slowly varying. 
However, we will see that if the parameter characterizing the strength of the potential, $\Phi_{0}$, is large enough, the tagged
particle density can vary on a smaller length scale. 

Without the external potential, the tagged particle distribution is uniform and equal to $1/V$. We are primarily interested in
the non-linear response regime, \textit{i.e} we chose $\Phi_{0}$ such that the tagged particle distribution is strongly
non-uniform. Specifically, we chose $\Phi_{0}=0.1$ for $T_a=0.01$ and $\Phi_{0}=1.0$ and $10.0$
for $T_a=1.0$. 

To improve the statistics we applied the external potential to $1.0\%$ of the particles for $T_a=0.01$ and 
$0.2\%$ of the particles for $T_a=1.0$. We note that while selecting the percentage of particles to which external
potential is applied one has to make sure that these particles are dilute enough in the whole system to be non-interacting.
To check for this we calculated steady state structure factors for the particles on which the external potential acts, in the 
plane perpendicular to the direction of the external force and confirmed that with these percentages the particles were dilute enough.
The fact that a smaller percentage is necessary to fulfill this condition at $T_a=1.0$ agrees with the insight from 
previously investigated potentials of the mean-force \cite{Berthier2017} that AOUPs at $T_a=1.0$ are effectively
larger than those at $T_a=0.01$. 

To further improve the statistics, we applied the external potential to different particles along each Cartesian coordinate 
within the same system. The distributions presented are averaged over all three Cartesian coordinates.

We evaluated tagged particle distributions due to external potential (\ref{eq:ExtPotential}). We fitted Boltzmann distributions 
(\ref{eq:TaggedProbDens}) to the numerical results treating the effective temperature as a fit parameter \cite{commentfit}. 
The resulting values are shown in the following as $T^\text{fit}$.

\subsubsection{Evaluation of the Einstein temperature}
	\label{subsubsec:T_E}

The Einstein effective temperature is defined as the ratio of the self-diffusion and tagged particle mobility coefficients  
\cite{About2004,Pottier2005},
\begin{equation}\label{TEdef} 
T^\text{E} = D/\mu.
\end{equation}

To calculate the self-diffusion coefficient we simulate unperturbed 
systems of AOUPs described before and use the standard
relation,
\begin{equation}\label{eq:Dcalc}
D = (2 d N)^{-1}\lim_{t\to\infty} \frac{1}{t} \sum_i \left<\left(\mathbf{r}_{i}(t)-\mathbf{r}_{i}(0)\right)^{2}\right>
\end{equation}
where $d$ is the dimensionality of the system. 

To evaluate the tagged particle mobility coefficient for our out-of-equilibrium systems we use the approach presented in 
Ref.~\cite{Szamel2017}, which involves the application of Malliavin weights. We define the mobility coefficient in terms
of a time-dependent response function $\chi(t)$ \cite{Szamel2017},
\begin{equation}\label{eq:mucalc}
\mu = \lim_{t\to\infty} \frac{1}{t} \chi(t).
\end{equation}
In turn, the response function is calculated through averages involving weighting functions~\cite{Szamel2017},
\begin{multline}\label{eq:ResponseFnx}
\chi(t) = (Nd)^{-1} \sum_{\alpha,i} \left[ \left<\alpha_{i}(t)\left(q_{i\alpha}(t)-q_{i\alpha}(0)\right)\right>
\right. \\ \left.
+\tau_p \left<\dot{\alpha}_{i}(t)\left(q_{i\alpha}(t)-q_{i\alpha}(0)\right)\right>\right],
\end{multline} 
where $\alpha=x, y, z$ and the 
weighting function $q_{i\alpha}(t)$ obeys the equation of motion,
\begin{equation}
\dot{q}_{i\alpha} = \mu_0^2/\left(2 T_{\text{a}}\right)^{-1} \eta_{i\alpha},
\end{equation}
where $\eta_{i\alpha}$ is the $\alpha$ component of the Gaussian noise acting on particle $i$.

\subsection{Results}
	\label{subsec:results}

\begin{figure}
	\centering
	\includegraphics[scale=0.27]{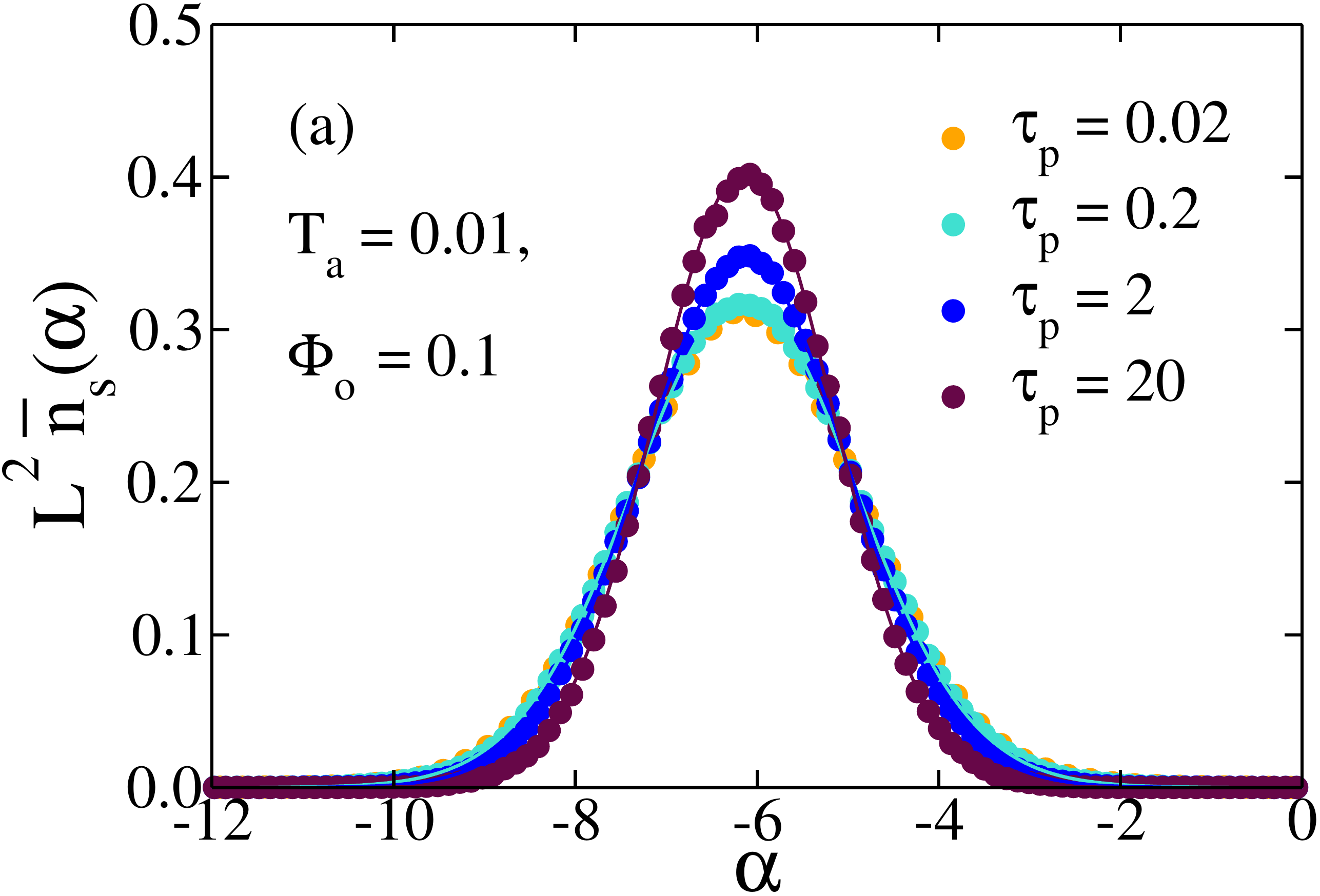}

	\vspace{5pt}	
	
	\includegraphics[scale=0.27]{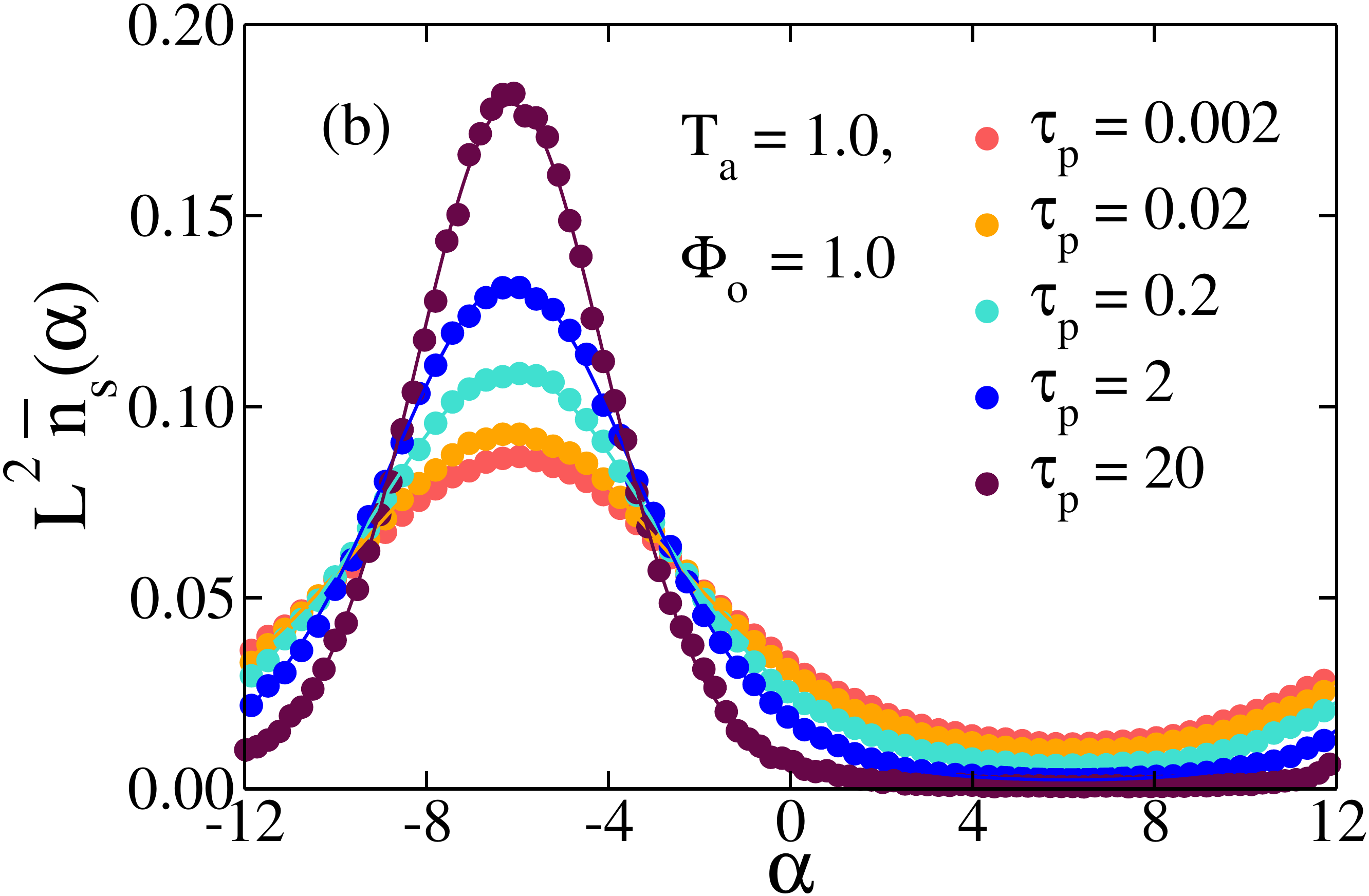}
	
	\vspace{5pt}
	
	\includegraphics[scale=0.27]{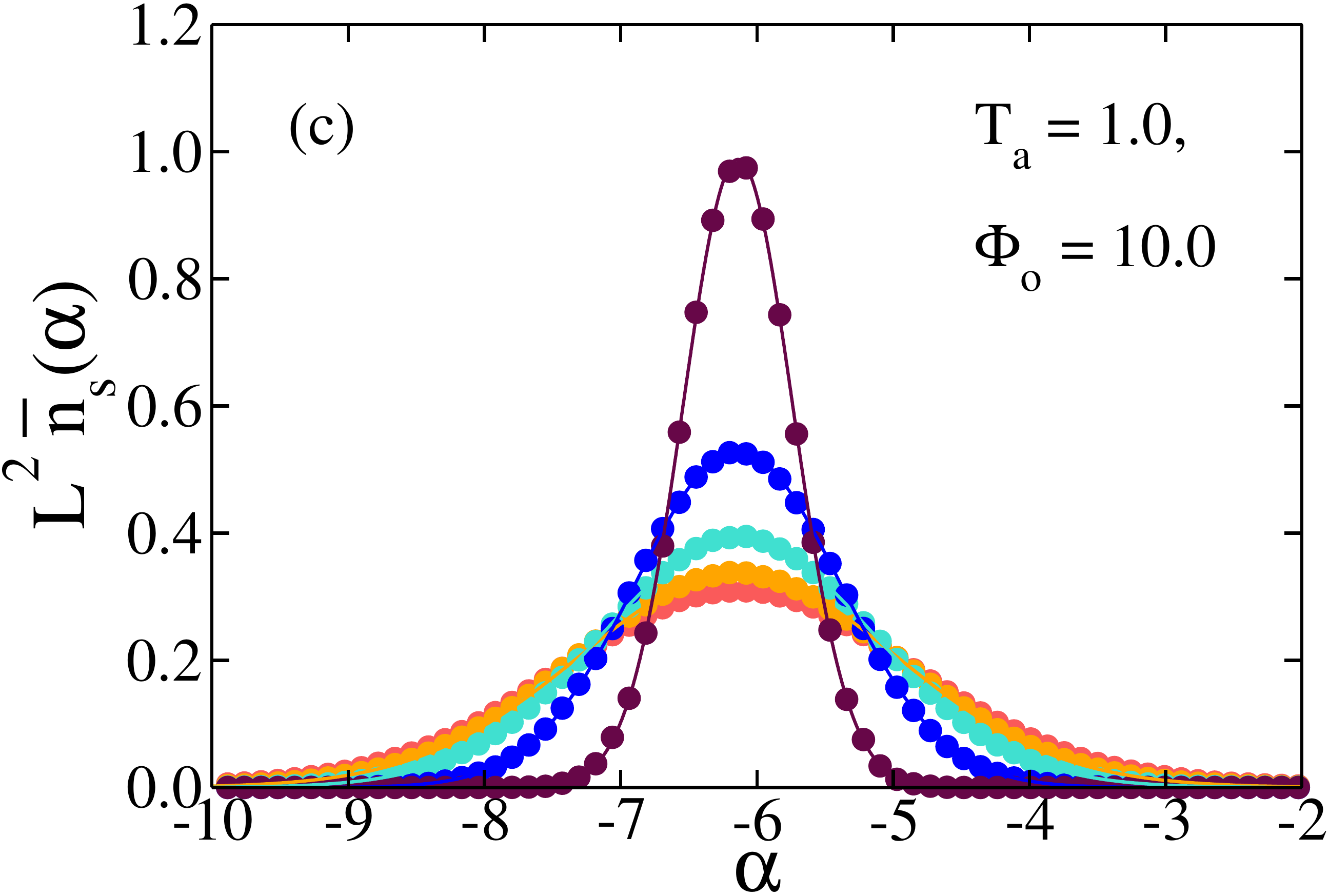}
	\caption{The tagged particle density distribution along the direction of the external potential,
$L^2 \bar{n}_s$, averaged over three different directions of the potential. 
(a) $T_a=0.01$, $\Phi_0=0.1$ and $\tau_p \in [0.02,20]$. (b) $T_a=1.0$, $\Phi_0=1.0$ and $\tau_p \in [0.002,20]$. 
(c) $T_a=1.0$, $\Phi_0=1.0$ and $\tau_p \in [0.002,20]$. Solid lines indicate Boltzmann distributions fitted to the data. The 
unperturbed distribution would be $L^2\times V^{-1}=L^{-1}=0.0407 $.}
	\label{fig:ns}
\end{figure}

For the external potential $\Phi$ varying along $\alpha$ axis, $\alpha=x,y,z$, the tagged particle distribution varies along 
the same direction and is uniform along the remaining two directions. In Fig.~\ref{fig:ns} we show tagged particle 
distributions averaged over the three directions of the perturbation. More precisely, we show $L^2 \bar{n}_s(\alpha)$ where 
$L=24.56$ is the box size and the over-bar denotes averaging over three directions of the perturbation. We use the standard
normalization condition $\int_V d\mathbf{r} \bar{n}_s(\mathbf{r}) = 1$. The distributions shown in Fig.~\ref{fig:ns} are
significantly different from the tagged particle density in the absence of the external potential, when $L^2 V^{-1} = 1/L =0.0407$.

The important qualitative information that can be obtained from a quick look at Fig.~\ref{fig:ns} is that the tagged particle
densities depend strongly on the persistence time of the self-propulsion. 

\begin{figure}
	\centering
	\includegraphics[scale=0.32]{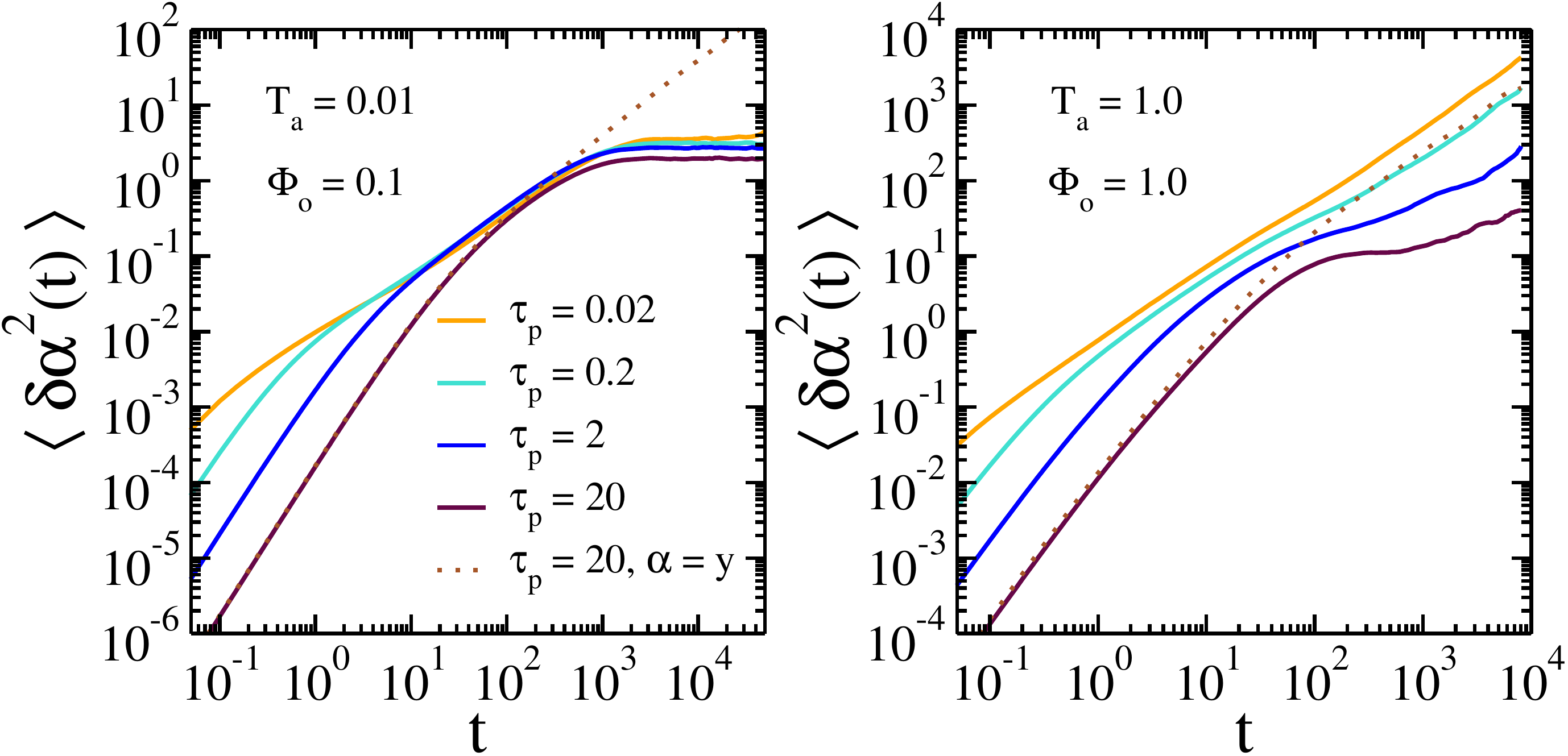}					
	\caption{Tagged particle mean squared displacement along the direction of the external potential. 
(a) $T_a=0.01$, $\Phi_0=0.1$ and $\tau_p \in [0.02,20]$.
Strong external potential leads to a localization of the tagged particle on the time scale of the simulation. 
(b) $T_a=1.0$, $\Phi_0=1.0$ and $\tau_p \in [0.02,20]$. Weaker external potential slows down the tagged particle motion
but does not localize it on the time scale of the simulation. Dashed lines show tagged particle mean squared displacement
in the direction perpendicular to the external potential for $\tau_p=20$. The motion in the perpendicular direction is
unperturbed by the external potential.}
	\label{fig:MSDPerturbed}
\end{figure}

To further verify that we are in the non-linear response regime in Fig.~\ref{fig:MSDPerturbed}, we show the tagged particle 
mean squared displacements (MSDs) along the axis of the external potential (solid curves) compared to the mean squared displacement
along an axis perpendicular to that of the external potential (dashed curves, for clarity shown for the longest persistence time only).
We see that the MSDs along the axis of the external potential are significantly different from those along a perpendicular axis. 
In fact, for $T_a=0.01$ and $\Phi_0=0.1$ the tagged particle is localized at the external potential minimum on the time scale 
of the simulation.

\begin{figure}
	\centering
	\includegraphics[scale=0.27]{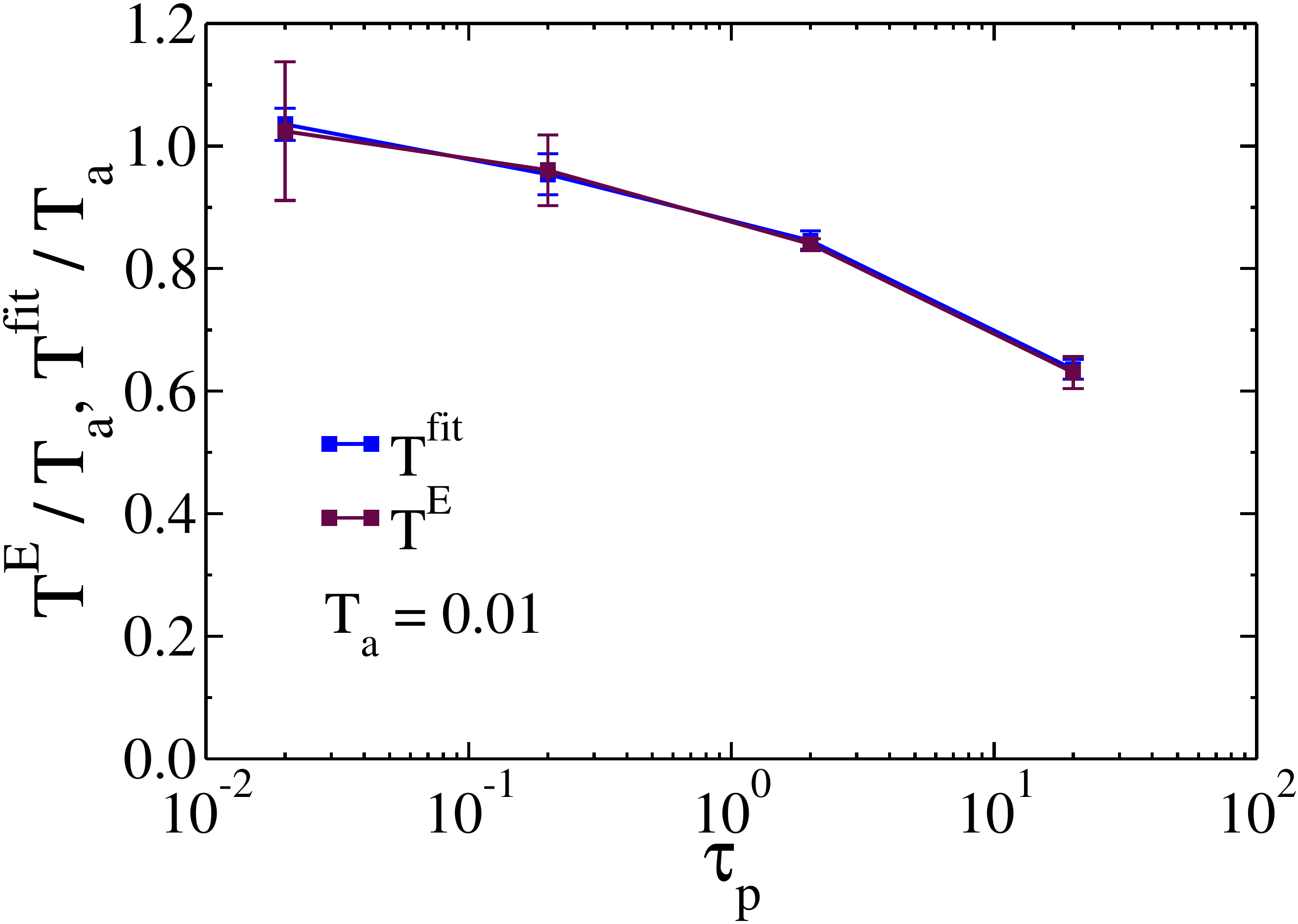}
	
	\vspace{5pt}
		
	\includegraphics[scale=0.27]{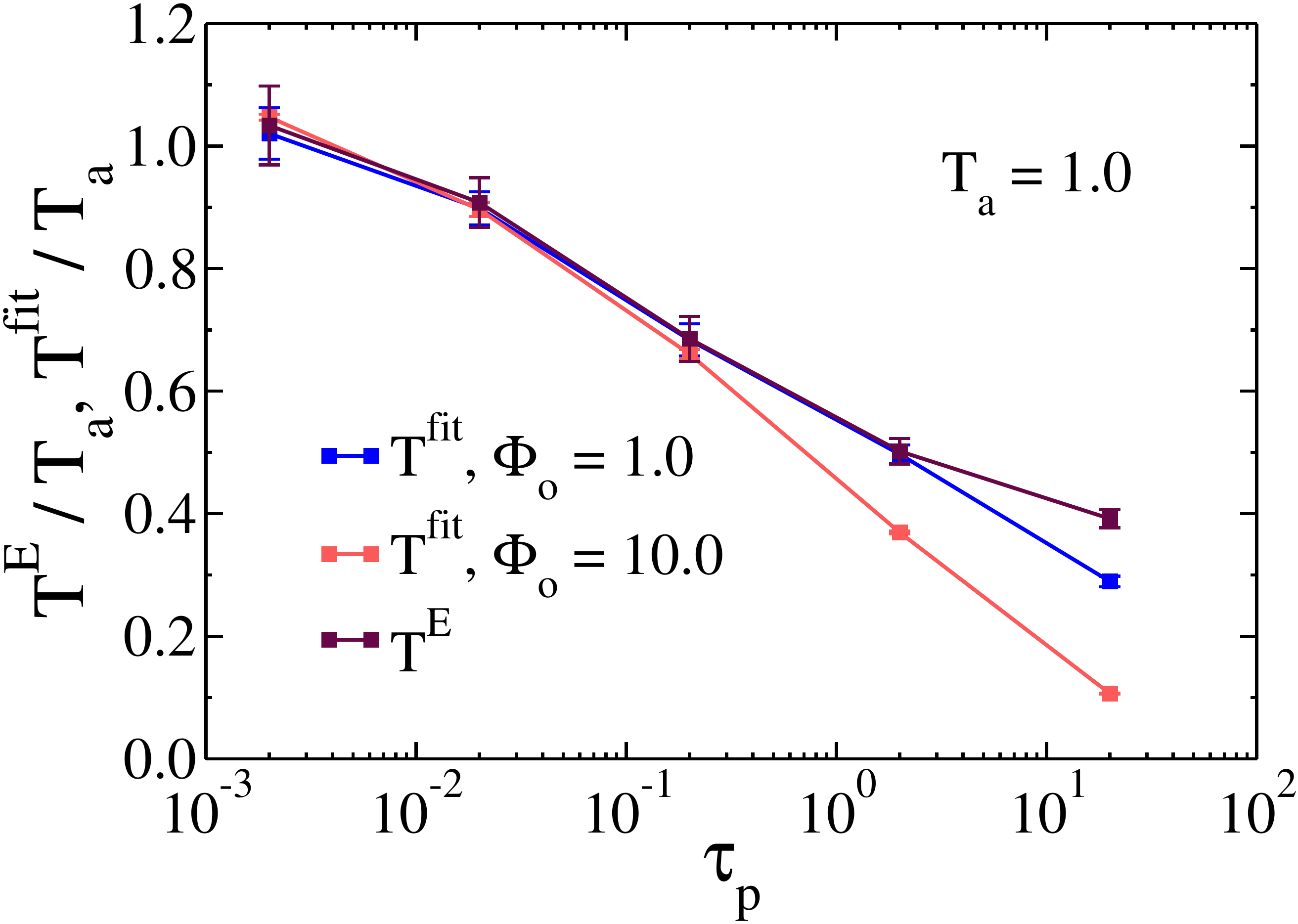}
			
	\caption{Comparison of the temperatures obtained from fitting Boltzmann distributions to tagged particle density
distributions and the Einstein relation effective temperatures. (a) $T_a=0.01$, $\Phi_0=0.1$ and $\tau_p \in [0.02,20]$.
(b) $T_a=1.0$, $\Phi_0=1.0$, and $\Phi_0=10.0$, and $\tau_p \in [0.002,20]$. All temperatures are shown relative to the
active temperature $T_a$.}
	\label{fig:EffTemp}
\end{figure}

The tagged particle density distributions shown in Fig.~\ref{fig:ns} can be fitted very well to the Boltzmann distribution 
$\propto \exp(-\Phi(\mathbf{r})/T^\text{fit})$ using $T^\text{fit}$ as the fit parameter \cite{commentfit}.
The resulting values $T^\text{fit}$ are shown in Fig.~\ref{fig:EffTemp}. We observe that the fitted temperatures decrease
with increasing persistence time, which could have been anticipated from the persistence time dependence of the tagged
particle densities. 

To verify our theory presented in Sec.~\ref{sec:theory} we need to check whether temperatures obtained from the fits, $T^\text{fit}$,
are the same as the Einstein temperatures obtained from the ratios of the self-diffusion and tagged particle mobility
coefficients. Even before calculating the latter temperatures we can infer from Fig.~\ref{fig:EffTemp} that the theory does not work
for the two longest persistence times for $T_a=1.0$. The reason is that the Einstein temperature describes an unperturbed system and thus
does not depend on $\Phi_0$ whereas for  the two longest persistence times for $T_a=1.0$ the temperatures obtained from the fits
depend on $\Phi_0$. We will return to this issue at the end of this section.

\begin{figure}
	\centering
	\includegraphics[scale=0.27]{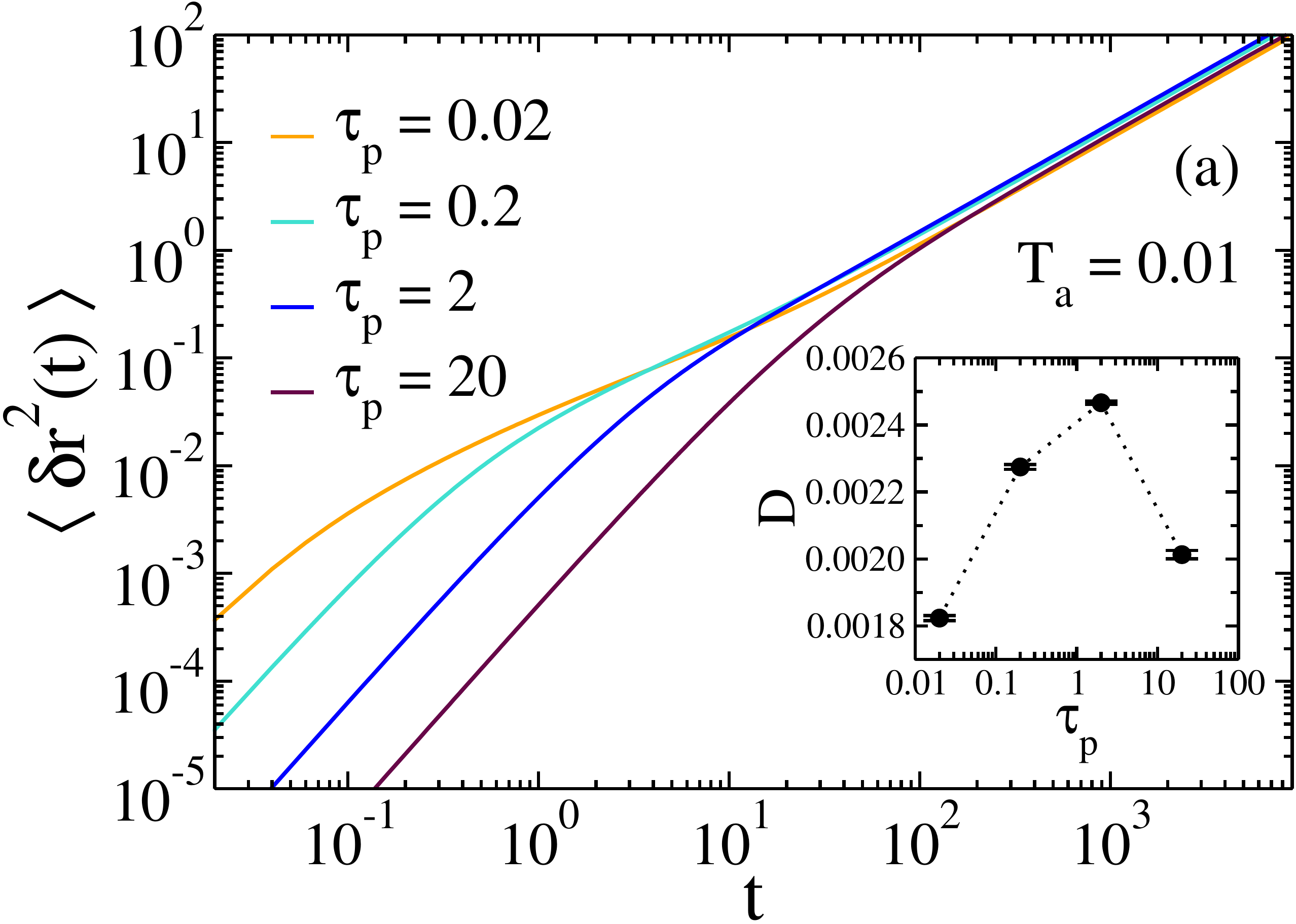}

	\vspace{5pt}

	\includegraphics[scale=0.27]{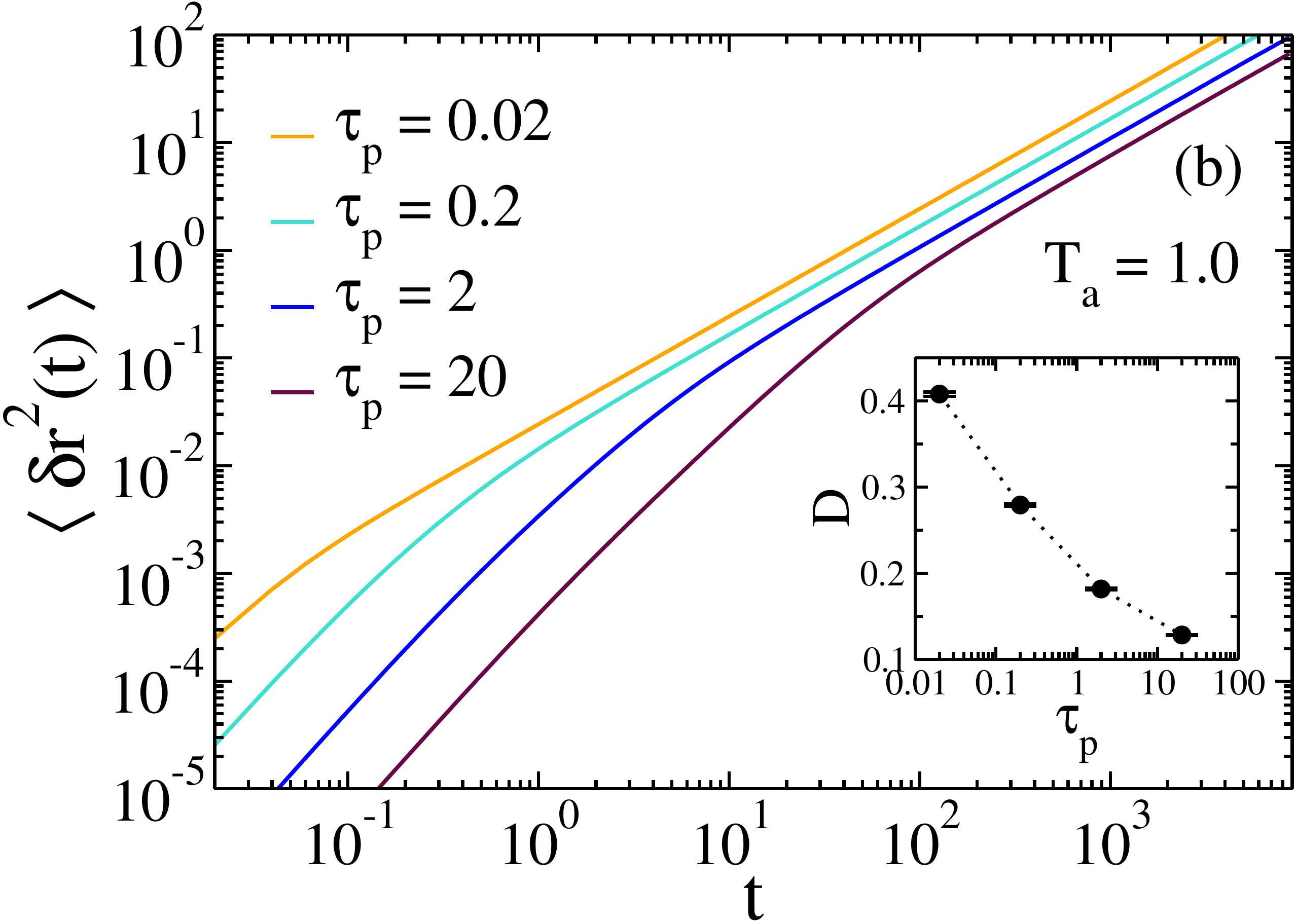}	
		
	\caption{Mean squared displacement in unperturbed systems. (a) $T_a=0.01$ and $\tau_p \in [0.02,20]$.
(b) $T_a=1.0$ and $\tau_p \in [0.02,20]$. Insets: persistence time dependence of the self-diffusion coefficient. 
$D$ is a non-monotonic function of $\tau_p$ at $T_a=0.01$ and decreases monotonically with increasing $\tau_p$ at $T_a=1.0$.}
	\label{fig:DiffusionConst}
\end{figure}

In Figs.~\ref{fig:DiffusionConst}(a-b) we show the MSDs $\left< \delta r^{2}(t)\right>$, where 
$\delta r^{2}(t) = \left(\mathbf{r}_1(t)-\mathbf{r}_1(0)\right)^2$, for unperturbed systems. The self-diffusion 
coefficients are calculated from these MSDs according to Eq. (\ref{eq:Dcalc}) and are presented in the insets. As anticipated and
in agreement with earlier investigations \cite{FlennerSzamel2020,Berthier2017}, we get two different behaviors of the 
self diffusion coefficient at two active temperatures investigated. 
For the lower active temperature, $T_a=0.01$, we observe a non-monotonic dependence of the self-diffusion coefficient on
the persistence time and for the higher active temperature, $T_a=1.0$, we observe that the self-diffusion coefficient decreases
monotonically with increasing persistence time. 

\begin{figure}[H]
	\centering
	\includegraphics[scale=0.27]{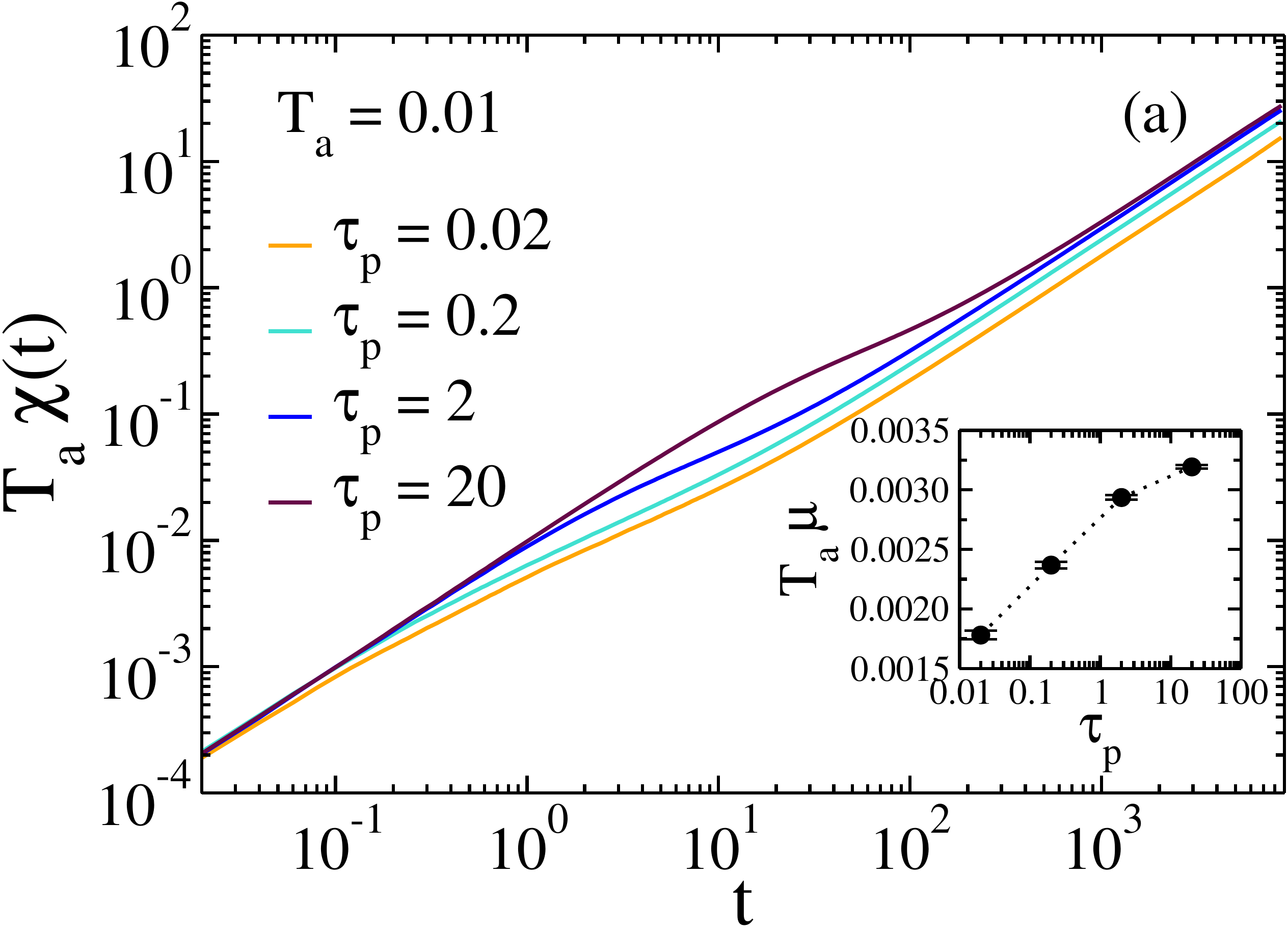}
	
	\vspace{5pt}
	
	\includegraphics[scale=0.27]{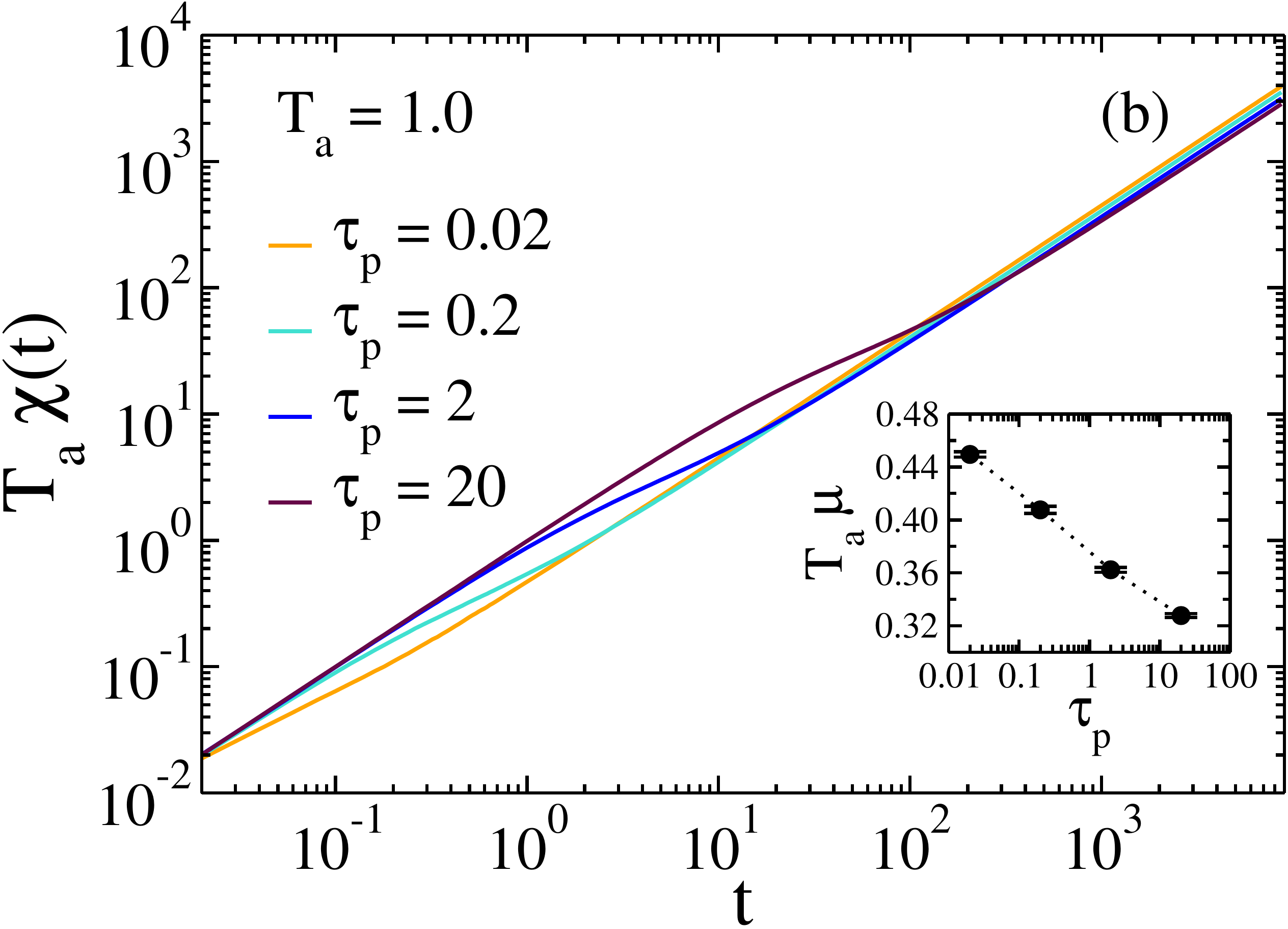}	
		
	\caption{Time dependent response function to a weak external potential in unperturbed systems 
calculated using Eq.~\ref{eq:ResponseFnx}. (a) $T_a=0.01$ and $\tau_p \in [0.02,20]$.
(b) $T_a=1.0$ and $\tau_p \in [0.02,20]$. Insets: persistence time dependence of the tagged particle mobility coefficient.
$T_a \mu$ decreases monotonically with increasing $\tau_p$ at both active temperatures.}
	\label{fig:MobilityConst}
\end{figure}

In Figs.~\ref{fig:MobilityConst}(a-b) we show the persistence time dependence of the time-dependent response function 
$\chi(t)$ \cite{Szamel2017} at the two active temperatures investigated.  The insets show the mobility coefficients calculated 
from the long time limit of $\chi(t)$ according to Eq. (\ref{eq:mucalc}).  
We note that at the lower active temperature, $T_{\text{a}}=0.01$, the mobility monotonically increases with increasing $\tau_{p}$, in
contrast to the non-monotonic behavior of the self-diffusion coefficient. At the higher active temperature, $T_{\text{a}}=1.0$, 
the mobility monotonically decreases with increasing persistence time, and thus exhibits the same $\tau_p$ dependence as the 
self-diffusion coefficient.

Comparing insets in Figs.~\ref{fig:DiffusionConst}(a-b) and in Figs.~\ref{fig:MobilityConst}(a-b) we can see that at the smallest 
persistence times $D\approx T_a \mu$. This behavior is expected since in the limit of the vanishing persistence time at constant 
active temperature, the present model active systems become equivalent to Brownian systems at temperature equal to the
active temperature, $T=T_a$. For a Brownian system the fluctuation-dissipation theorem holds and $D=T\mu$.

In Fig.~\ref{fig:EffTemp} we compare the Einstein temperatures $T^\text{E}$ defined as the ratios $D/\mu$ to the temperatures
obtained from fits to the Boltzmann distribution, $T^\text{fit}$. As mentioned in the previous paragraph, 
in the limit of small persistence 
times our active system becomes equivalent to the Brownian system and both $T^\text{fit}$ and $T^\text{E}$ become equal to 
the active temperature. With increasing persistence time, while keeping the active temperature constant, both $T^\text{fit}$ 
and $T^\text{E}$ decrease. We note that the decrease of the ratio of the Einstein temperature and the active temperature, 
$T^\text{E}/T_a$, was observed before \cite{Szamel2017,FlennerSzamel2020}. It contrasts with the increase 
of the ratio of the Einstein effective temperature to the bath temperature with 
increasing shear rate for colloidal suspensions under steady shear \cite{SzamelZhang2011}. 

For the lower active temperature, $T_a=0.01$, we observe a very good agreement between $T^\text{fit}$ and $T^\text{E}$ 
for all persistence times
investigated. In contrast, for the higher active temperature, $T_a=1.0$, we initially see a very good agreement between $T^\text{fit}$ 
and $T^\text{E}$ but then, for longer persistence times we observe that temperatures obtained from the fits deviate from the temperatures
from the Einstein relation. Notably, it happens first for $T^\text{fit}$ obtained for the more confining potential, 
$\Phi_0=10.0$, and then $T^\text{fit}$ obtained for the less confining potential, $\Phi_0=1.0$. 

We recall that our theoretical derivation in Sec.~\ref{sec:theory} relied upon the assumption that the spatial variation of the 
potential and of the tagged particle density occurs on the longest relevant length scale. On the other hand, we know that 
with increasing persistence time systems of self-propelled particles may undergo a motility-induced phase separation and
that upon approaching such a transition they can exhibit long-range density fluctuations. To investigate the existence of such 
fluctuations we evaluated steady state structure factors of the unperturbed systems. 

\begin{figure}
	\centering
	\includegraphics[scale=0.27]{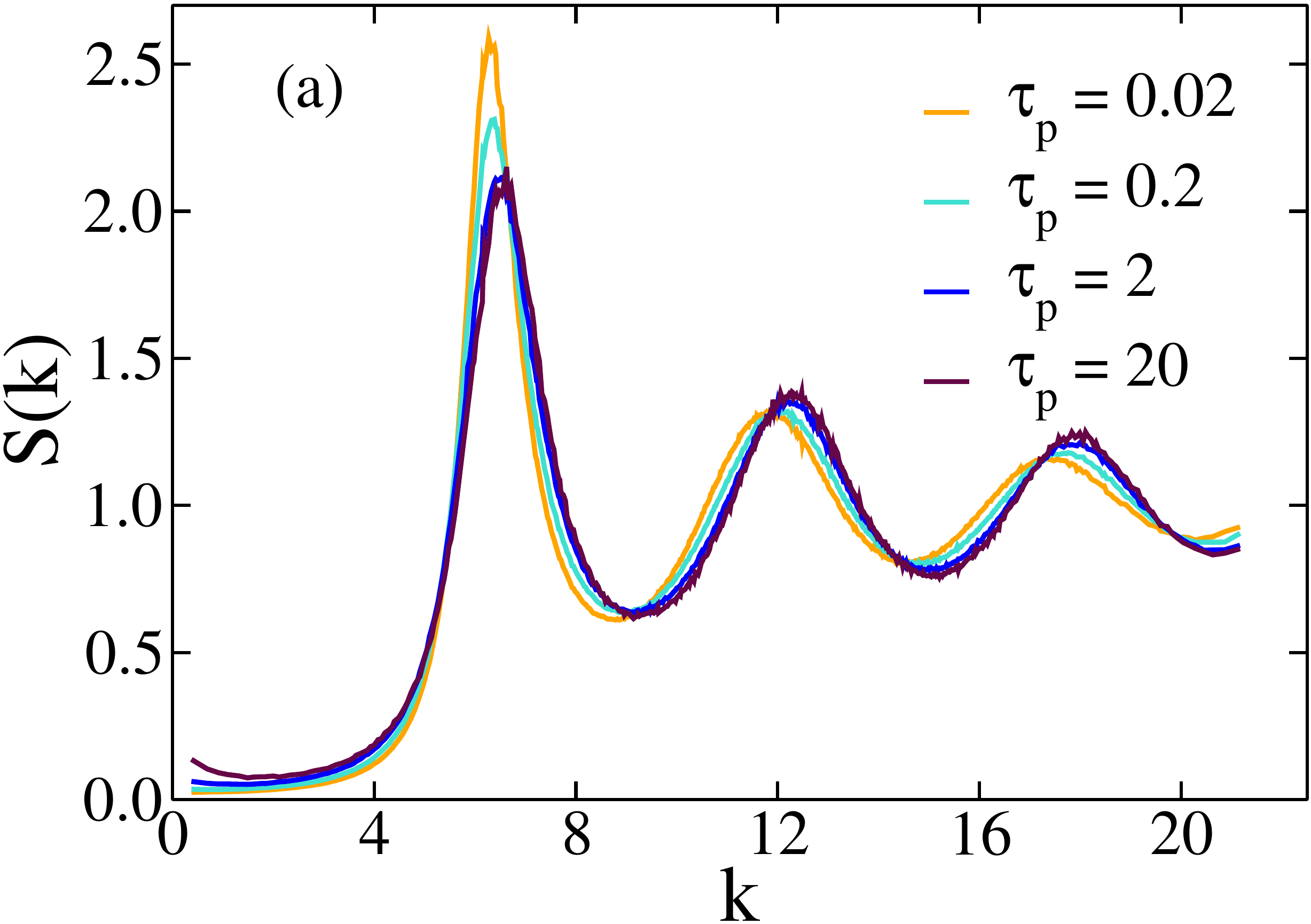}
	
	\vspace{5pt}
	
	\includegraphics[scale=0.27]{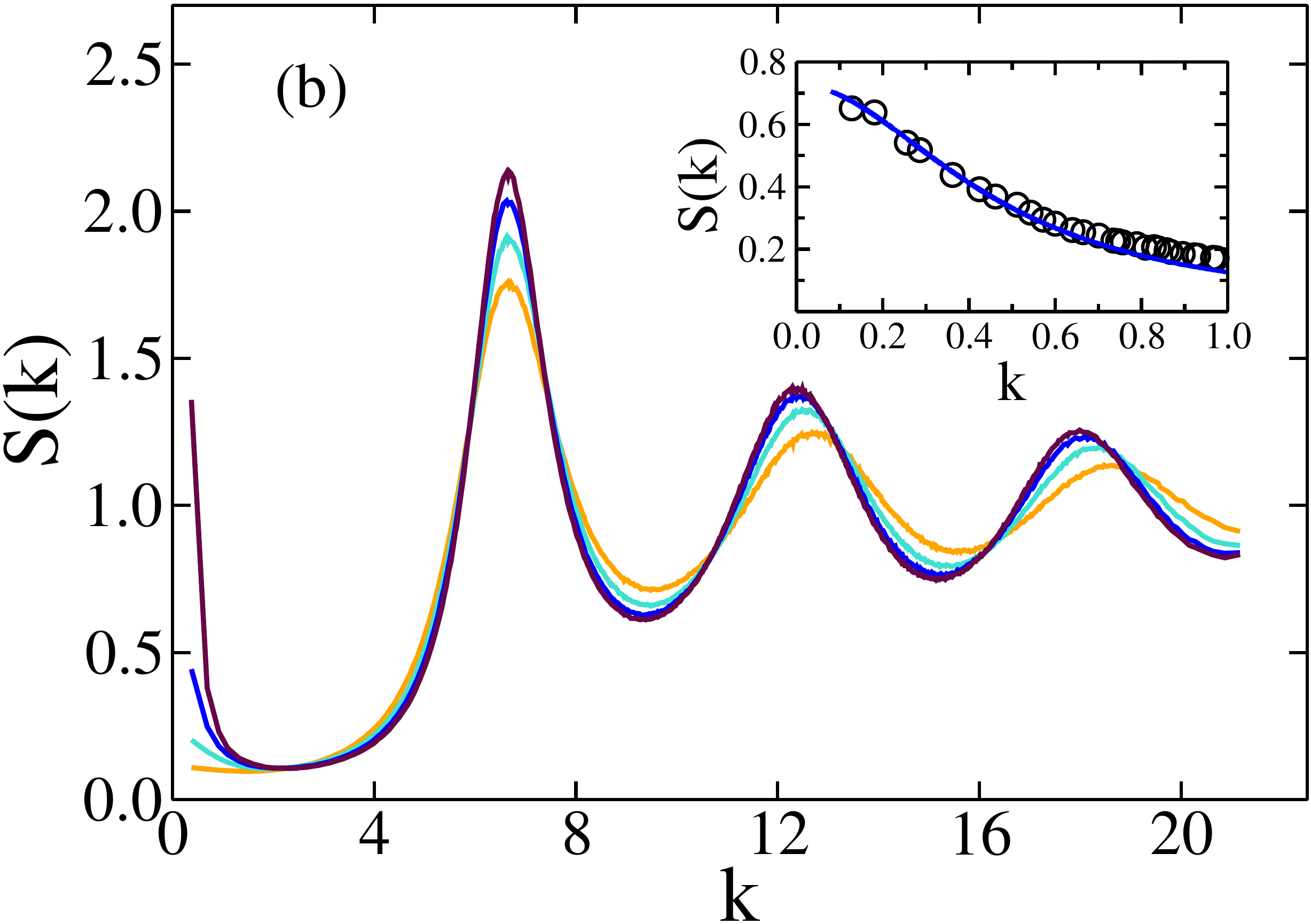}				
	\caption{Stationary state structure factors of unperturbed systems. (a) $T_a=0.01$ and $\tau_p \in [0.02,20]$.
(b) $T_a=1.0$ and $\tau_p \in [0.02,20]$. The upturn at small wavevectors indicates increasing correlation length.
Inset in (b): small wavevector behavior of the structure factor calculated using $8\times 10^4$ particle system for 
$T_{\text{a}}=1.0$ and $\tau_{p}=2$. The solid line shows an Ornstein-Zernicke function, $f(k) = a/[1+(b k)^2]$, fitted to the data.}
	\label{fig:strFactor}
\end{figure}

In Fig.~\ref{fig:strFactor} we show steady state structure factors $S(k)$,
\begin{equation}
S(k)=1+\frac{1}{N} \left< \sum_{i=1}^{N} \sum_{j \neq i}^{N} \exp \left[-i \mathbf{k} \cdot\left(\mathbf{r}_{i}-\mathbf{r}_{j}\right)
\right]\right>,
	\label{eq:strFactor}
\end{equation}
for unperturbed systems at both active temperatures. We observe that for the lower active temperature, $T_a=0.01$, only a modest
increase of $S(k)$ is observed for small wavevectors at the longest persistence times. This suggests that at this active temperature 
and in the range of the persistence times investigated density correlations are relatively short-ranged. 

In contrast, for the higher active temperature, $T_a=1.0$, we observe a large small wavevector increase of $S(k)$ for the two
longest persistence times. This suggests that at this active temperature at these persistence times there are long-ranged
density fluctuations. To make this statement more quantitative we simulated a larger system consisting of $8\times 10^4$ particles
at $\tau_p=2.0$. The small wavevector behavior of the steady state structure factor for this system is shown in the inset to
Fig.~\ref{fig:strFactor}. To quantify the range of the density correlations we fitted the numerical results to the Ornstein-Zernicke form
$f(k) = a/[1+(bk)^2]$. We recall that parameter $b$ in the Ornstein-Zernicke fit is a measure of the density correlation length. 
We obtained $b=2.19$ which is perhaps moderate but is larger than the length on which the tagged particle density varies
for $\Phi_0=10.0$ at $T_a=1.0$, $\tau_p=2.0$. Thus, in hindsight, it is not surprising that our theory is not applicable for these
parameters.

\section{Conclusions}
	\label{sec:conclusions}

We derived an expression for the tagged particle density distribution in a slowly varying in space external potential
in a system of interacting athermal active particles. The tagged particle distribution has the Boltzmann functional form,
but the role of the temperature is played by the ratio of the self-diffusion and tagged particle mobility coefficients. 
We used computer simulations to verify the theoretical result. The theory works well if the characteristic length of the 
tagged particle density variation is the longest relevant length in the system. The theory is inapplicable if the characteristic length
of the density fluctuations is longer than the the characteristic length of the tagged particle density variation.

The ratio of the self-diffusion and tagged particle mobility coefficients has long been known as the Einstein temperature, 
one of several effective temperatures obtained for non-equilibrium systems from the fluctuation-dissipation ratios. 
Our result shows that the Einstein temperature determines the large spatial scale tagged particle density distribution 
beyond the linear response regime. This resembles earlier results that established that the Einstein temperature
plays similar role for a single Brownian particle in a tilted periodic potential and for a tagged particle in a colloidal 
suspension under steady shear flow. These three results obtained for very different systems 
suggest that the Einstein temperature may be generally relevant for
the large spatial scale tagged particle density distribution in any stationary non-equilibrium system in which the large scale
motion is diffusive.

\section*{Acknowledgments}

We gratefully acknowledge the support
of NSF Grant No.~CHE 1800282.

\end{document}